\documentclass[12pt,english,floatfix,superscriptaddress,aps,prd,preprint,showkeys,nofootinbib]{revtex4}
\usepackage{amsmath}
\usepackage{amssymb}
\usepackage{amsbsy}
\usepackage{amsfonts}
\usepackage{amsopn}
\usepackage{amstext}
\usepackage{graphicx}
\usepackage[english]{babel}
\usepackage{color}
\usepackage{slashed}
\usepackage{esint}
\usepackage[dvips]{epsfig}
\usepackage[dvips]{graphicx}
\usepackage{float}
\usepackage{units}
\usepackage{textcomp}
\usepackage{wasysym}
\usepackage{hyperref}
\usepackage{slashed}

\begin{document}

\title{Duality in $SIM(2)$ topologically massive models with $B\wedge F$ term}

\author{Fernando M. Belchior}
\email{belchior@fisica.ufc.br}
\affiliation{Departamento de Física, Universidade Federal do Ceará (UFC),\\ Campus do Pici, 60455-760, Fortaleza, Ceará, Brazil.}

\author{Roberto V. Maluf}
\email{r.v.maluf@fisica.ufc.br}
\affiliation{Departamento de Física, Universidade Federal do Ceará (UFC),\\ Campus do Pici, 60455-760, Fortaleza, Ceará, Brazil.}

\begin{abstract}
This paper aims to investigate the classical duality between the $SIM(2)$-Maxwell-Kalb-Ramond (MKR) theory and a self-dual non-gauge-invariant model. First, we establish the equivalence in the free-field case using two complementary methods: a direct comparison of the equations of motion and the master Lagrangian approach. In both methodologies, we verify that the classical correspondence between the MKR model and self-dual fields exhibits modifications induced by very special relativity (VSR). Moreover, we employ the master Lagrangian approach to examine the duality when the self-dual model is minimally coupled to fermionic matter. We show that the resulting MKR model contains Thirring-like interactions modified by nonlocal contributions arising from VSR.
\end{abstract}

\maketitle

%%%%%%%%%%%%%%%%%%%%%%%%%%%%%%%%%%%%%%%%%%%%%%%%%%%%%%%%%%%%%%%%%%%%
\section{Introduction}

%The phenomenon of mass generation is a central issue in the context of fundamental interactions, with the Higgs mechanism providing the primary explanation for particle masses. Another important challenge is the prediction of neutrino masses, which are known to be exceedingly small. In this context, violations of Lorentz symmetry can be invoked to address this issue \cite{Cohen:2006ir, Kostelecky:2003cr}.

The possibility of Lorentz symmetry violation as a signal of new physics has motivated the development of the Standard Model Extension (SME), a comprehensive generalization of the Standard Model that includes Lorentz- and CPT-violating operators \cite{Colladay:1996iz, Colladay:1998fq}. Such terms may arise from an underlying fundamental theory. More recently, a framework known as very special relativity (VSR) was proposed by Cohen and Glashow \cite{Cohen:2006ir} as an alternative approach to investigating Lorentz symmetry breaking (LSB) and the origin of the neutrino mass. In this framework, the full Lorentz group is reduced to two subgroups, namely the homothety group $HOM(2)$ and the similitude group $SIM(2)$ \cite{Cohen:2006ky}. VSR allows the introduction of neutrino masses without requiring additional particles. Beyond this context, several physical systems exhibiting VSR effects have been studied, including gauge theories \cite{Alfaro:2013uva, Alfaro:2019koq, Bufalo:2019qot, Shah:2018fms, Cheon:2009zx}, the Elko field \cite{Ahluwalia:2010zn}, and gravitational models \cite{Gibbons:2007iu, Kouretsis:2008ha}.

On the other hand, considerable attention has been devoted to the study of dualities. This theme has played a central role in exploring the interface between high-energy and condensed matter physics, particularly in the investigation of nonperturbative aspects of low-dimensional field theories \cite{Burgess:1993np}. In this context, bosonization stands out as a powerful tool, leading to the emergence of novel particle–vortex dualities in $2+1$ dimensions and in topological insulators \cite{Marino:1990yi, Fradkin:1994tt, Metlitski:2015eka, Hernaski:2018jsy, Karch:2016sxi, Santos:2019dlr, Mross:2017gny, Hsiao:2023hzd, Aharony:2012nh}. In string theory, Maldacena proposed a seminal conjecture, the so-called $AdS/CFT$ correspondence, which relates string theory in higher-dimensional anti–de Sitter spacetime to a conformal field theory (CFT) on its boundary \cite{Maldacena:1997re}. Additionally, $T$- and $S$-dualities, which also arise in string theory \cite{Strominger:1996it, Polchinski:2014mva}, further enrich this subject.

In the context of topologically massive theories in $2+1$ dimensions, the seminal work of Deser and Jackiw established the equivalence between the Maxwell–Chern–Simons (MCS) and self-dual (SD) models \cite{Deser:1984kw}. Subsequently, several studies exploring extensions of this duality, including couplings to matter \cite{Ilha:2001he, Anacleto:2000ea, Dalmazi:2006yv}, noncommutative generalizations \cite{Gomes:2008pi}, Lorentz symmetry breaking \cite{BaetaScarpelli:2014jfl, Furtado:2008gs, Marques:2022hhl}, and supersymmetric extensions \cite{Ferrari:2008he, Ferrari:2006vy} were published.

The equivalence between topological models is usually examined using two different methods: gauge embedding \cite{Anacleto:2001rp} and the master action \cite{Dalmazi:2008zh}. In the first approach, gauge symmetry is revealed through an embedding procedure applied to the SD model, leading to the MCS theory \cite{Marques:2022hhl}. The second method, known as the master Lagrangian approach, constructs a first-order Lagrangian that interpolates between the two theories. Unlike the gauge-embedding method, the master Lagrangian preserves gauge symmetry, allowing the proof of duality at the quantum level.

Besides the free-field case, it is essential to verify the duality when matter fields are included. In particular, the Chern–Simons theory minimally coupled to fermions has played a pivotal role in the study of three-dimensional quantum field theories with applications to condensed matter systems \cite{Seiberg:2016gmd}. As shown in Ref. \cite{Gomes:1997mf}, when the self-dual model is minimally coupled to Dirac fermions, a Thirring-like term must be added to the MCS Lagrangian in order to preserve the equivalence between the two theories. In this sense, it has been shown that a conjectured fermion–boson correspondence leads to the emergence of a nontrivial web of dualities, with interesting connections to topologically ordered systems, as investigated in Refs. \cite{Karch:2016sxi,Hernaski:2018jsy}. Moreover, the appearance of Thirring-like interactions is crucial for describing interacting electrons confined to lower spatial dimensions. In this framework, duality symmetries and the bosonization procedure can be employed to provide exact solutions to nonperturbative problems \cite{Fradkin:1994tt}.

The topological duality can be extended to $3+1$ dimensions \cite{Menezes:2002nj, Maluf:2020fch, Gama:2022nue} through the $B \wedge F$ term, where the topologically massive theory is constructed from a Lagrangian density involving a vector field $A_\mu$ and an antisymmetric rank-two tensor field $B_{\mu\nu}$, known as the Kalb--Ramond field \cite{Kalb:1974yc, Maluf:2023gbf}. 

Following Ref.~\cite{Belchior:2024eyz}, in which the duality between the SD model and the MCS theory was investigated within the VSR framework, we are interested in examining the corresponding duality in $3+1$ dimensions with the implementation of a $SIM(2)$-invariant $B \wedge F$ term. As discussed previously, the $SIM(2)$-VSR gauge symmetry gives rise to novel nonlocal interaction terms \cite{Alfaro:2019koq, Maluf:2023gbf}. In this context, our main goal is to analyze how these VSR-induced nonlocal terms affect the duality between theories involving a vector field and the Kalb--Ramond field.

This paper is organized as follows. In Sec.~\ref{s2}, we present the self-dual model with a $SIM(2)$ $B \wedge F$ term, along with the corresponding equations of motion. In Sec.~\ref{s3}, we introduce the $SIM(2)$ topologically massive theory with a $B \wedge F$ term. In Sec.~\ref{s4}, we establish the classical duality using the master Lagrangian approach. The duality in the presence of fermionic matter couplings is analyzed in Sec.~\ref{s5}. Finally, our conclusions and perspectives for future work are presented in Sec.~\ref{s6}.

%%%%%%%%%%%%%%%%%%%%%%%%%%%%%%%%%%%%%%%%%%%%%%%%%%%%%%%%%%%%%%%%%%%%%%%%%%%%%%%%%
\section{$B\wedge F$ Self-Dual in VSR}\label{s2}

In the context of $3+1$-dimensional theories, there exists a topologically massive model involving a vector field $A_\mu$ and an antisymmetric tensor field $B_{\mu\nu}$. This model is commonly referred to as the self-dual $B \wedge F$ theory ($SD_{B \wedge F}$) and is described by the following first-order Lagrangian density \cite{Maluf:2020fch}.
\begin{align}
\mathcal{L}=\frac{m^2}{2}A_\mu A^\mu -\frac{1}{4}B_{\mu\nu} B^{\mu\nu}+\frac{\theta}{2} \epsilon^{\mu\nu\alpha\beta}\partial_\mu A_\nu  B_{\alpha\beta}.
\end{align}
It is straightforward to verify that the Lagrangian above does not exhibit gauge invariance. In addition, we may construct a $SIM(2)$ version of the $SD_{B\wedge F}$ model, described by the Lagrangian density:
\begin{align}\label{sfvsr}
\mathcal{L}_{SD}=\frac{m^2}{2}A_\mu A^\mu -\frac{1}{4}B_{\mu\nu} B^{\mu\nu}+\frac{\theta}{4} \epsilon^{\mu\nu\alpha\beta}\widetilde{\partial}_\mu A_\nu  B_{\alpha\beta}+\frac{\theta}{4} \epsilon^{\mu\nu\alpha\beta} A_\mu \overset{\nsim}{ \partial_\nu}B_{\alpha\beta},
\end{align}
where we have defined the following $SIM(2)$ derivatives 
\begin{align}
\widetilde{\partial}_\mu=\partial_\mu+\frac{m_{A}^2}{2}\frac{n_\mu}{(n\cdot\partial)},\\
\overset{\nsim}{ \partial}_\mu=\partial_\mu+\frac{m_{B}^2}{2}\frac{n_\mu}{(n\cdot\partial)}.
\end{align}
Here, $m_A$ and $m_B$ denote the VSR-induced masses associated with the $A_\mu$
and the $B_{\mu\nu}$ fields, respectively. The constant null vector $n_\mu = (1, 0, 0, 1)$ selects a preferred direction in spacetime. By varying the action corresponding to $\mathcal{L}_{SD}$, we obtain the equations of motion for the $A_\mu$ and $B_{\mu\nu}$ fields, namely,
\begin{align}
\label{ebf1} m^2 A^\mu+\frac{\theta}{2}\epsilon^{\mu\nu\alpha\beta}\partial_\nu B_{\alpha\beta}+\frac{\theta M^2}{8}\epsilon^{\mu\nu\alpha\beta}\frac{n_\nu}{(n\cdot\partial)}B_{\alpha\beta}=0,\\
\label{ebf2}    -\frac{1}{2}B^{\mu\nu}+\frac{\theta}{2}\epsilon^{\mu\nu\alpha\beta}\partial_\alpha A_\beta +\frac{\theta M^2}{8}\epsilon^{\mu\nu\alpha\beta}\frac{n_\alpha}{(n\cdot\partial)}A_\beta=0,
\end{align}
where $M^2=m_{A}^2+m_{B}^2$. Besides, the fields satisfy the following constraints
\begin{align}
    m^2\partial_\mu A^\mu+\frac{M^2}{4}\frac{n_\mu}{(n\cdot\partial)} A^\mu=0,\\
     m^2\partial_\mu B^{\mu\nu}+\frac{M^2}{4}\frac{n_\mu}{(n\cdot\partial)}B^{\mu\nu}=0.
\end{align}
Taking these constraints into account, Eqs.~(\ref{ebf1}) and (\ref{ebf2}) can be written in the form of wave equations,
\begin{align}
    \bigg(\square+\frac{m^2}{\theta^2}+\frac{M^2}{2}\bigg)A_\mu=0,\\
    \bigg(\square+\frac{m^2}{\theta^2}+\frac{M^2}{2}\bigg)B_{\mu\nu}=0.
\end{align}
As can be seen, the first-order Lagrangian describes the propagation of a massive vector field, while the antisymmetric tensor field plays the role of an auxiliary field \cite{Maluf:2020fch}. The next step is to construct a $SIM(2)$ gauge-invariant Lagrangian with the $B\wedge F$ term, which will be presented in the next section.

%%%%%%%%%%%%%%%%%%%%%%%%%%%%%%%%%%%%%%%%%%%%%%%%%%%%%%%%%%%%%%%%%%%%%%%%%%
\section{$SIM(2)$ Maxwell-Kalb-Ramond model}\label{s3}
Let us now consider a second-order theory with a topologically massive $SIM(2)$ $B\wedge F$ term, described by the Lagrangian density
\begin{align}\label{MKRvsr}
\mathcal{L}_{MKR}=\frac{\theta^2}{12m^2}\overset{\nsim}{H}_{\lambda\mu\nu} \overset{\nsim}{H} ^{\lambda\mu\nu}-\frac{\theta^2}{4}\widetilde{F}_{\mu\nu}\widetilde{F}^{\mu\nu}-\frac{\theta}{4} \epsilon^{\mu\nu\alpha\beta}\widetilde{\partial}_\mu A_\nu  B_{\alpha\beta}-\frac{\theta}{4} \epsilon^{\mu\nu\alpha\beta} A_\mu \overset{\nsim}{ \partial_\nu}B_{\alpha\beta},  
\end{align}
where $\widetilde{F}_{\mu\nu}=\widetilde{\partial}_\mu A_\nu-\widetilde{\partial}_\nu A_\mu$ 
denotes the field strength associated with the vector field $A_\mu$, and $\overset{\nsim}{H}_{\lambda\mu\nu}=\overset{\nsim}{\partial}_\lambda B_{\mu\nu}+\overset{\nsim}{\partial}_\mu B_{\nu\lambda}+\overset{\nsim}{\partial}_\nu B_{\lambda\mu}$ is the corresponding field strength of the Kalb-Ramond field $B_{\mu\nu}$. The action associated with Eq. (\ref{MKRvsr}) is invariant under the transformations
\begin{align}
    A_\mu &\rightarrow A _\mu + \widetilde{\partial}_\mu\,\Lambda,\\
    B_{\mu\nu} \rightarrow & \, B_{\mu\nu}+ \overset{\nsim}{\partial}_\mu\Sigma_\nu-\overset{\nsim}{\partial}_\nu\Sigma_\mu,
\end{align}
where $\Lambda$ and $\Sigma$ are arbitrary gauge parameters. The gauge parameter $\Sigma$ admits a subsidiary gauge transformation of the form $\Sigma_\mu \rightarrow \Sigma _\mu + \overset{\nsim}{\partial}_\mu\,\alpha$, where $\alpha$ is an additional gauge parameter. 
Furthermore, the $SIM(2)$ derivatives have been defined previously. Accordingly, the field strengths $\widetilde{F}_{\mu\nu}$ and $\overset{\nsim}{H}_{\lambda\mu\nu}$ can be recast using the $SIM(2)$ derivatives as \cite{Maluf:2023gbf}
\begin{align}
\widetilde{F}_{\mu\nu}=F_{\mu\nu}+\frac{m_{A}^2}{2}\bigg[\frac{n_\mu n^\lambda}{(n\cdot\partial)^2}F_{\lambda\nu}-\frac{n_\nu n^\lambda}{(n\cdot\partial)^2}F_{\lambda\mu}\bigg],   
\end{align}
and
\begin{align}
\overset{\nsim}{H}_{\lambda\mu\nu}=H_{\lambda\mu\nu}+\frac{m_{B}^2}{2}\bigg[\frac{n_\lambda n^\rho}{(n\cdot\partial)^2}H_{\rho\mu\nu}+\frac{n_\mu n^\rho}{(n\cdot\partial)^2}H_{\rho\nu\lambda}+\frac{n_\nu n^\rho}{(n\cdot\partial)^2}H_{\rho\lambda\mu}\bigg].  
\end{align}

With these definitions, the Lagrangian in Eq.~(\ref{MKRvsr}) can be rewritten in ordinary spacetime as follows:
\begin{align}\label{Ltm}
\mathcal{L}_{MKR}&=\frac{\theta^2}{m^2}\bigg[\frac{1}{12}H_{\lambda\mu\nu}H^{\lambda\mu\nu}+\frac{1}{2}m_{B}^2 n_\lambda\Big(\frac{1}{n\cdot\partial}H^{\lambda\mu\nu}\Big)n_\rho\Big(\frac{1}{n\cdot\partial}H^\rho\ _{\mu\nu}\Big)\bigg]\nonumber\\&+\theta^2\bigg[-\frac{1}{4} F_{\mu\nu}F^{\mu\nu}+\frac{1}{2}m_{A}^2 n_\mu\Big(\frac{1}{n\cdot\partial}F^{\mu\nu}\Big)n^\lambda\Big(\frac{1}{n\cdot\partial}F_{\lambda\nu}\Big)\bigg]\nonumber\\&-\frac{\theta}{8} \epsilon^{\mu\nu\alpha\beta}F_{\mu\nu}  B_{\alpha\beta}-\frac{\theta m_{B}^2}{8} \epsilon^{\mu\nu\alpha\beta}\frac{n_\mu n^\lambda}{(n\cdot\partial)^2}F_{\lambda\nu}  B_{\alpha\beta}\nonumber\\&-\frac{\theta}{12} \epsilon^{\mu\nu\alpha\beta} A_\mu H_{\nu\alpha\beta}-\frac{\theta}{8} \epsilon^{\mu\nu\alpha\beta} A_\mu \frac{n_\nu n^\lambda}{(n\cdot\partial)^2} H_{\lambda\alpha\beta}.  
\end{align}    

Thus, using Eq.~(\ref{Ltm}), we can derive the following equations of motion:
\begin{align}
\label{krm1}
\theta^2\bigg[\partial_\mu F^{\mu\nu}+m_{A}^2\frac{n_\mu}{(n\cdot\partial)}F^{\mu\nu}+\frac{m_{A}^2}{2}\frac{n^\nu n_\lambda}{(n\cdot\partial)^2}\partial_\mu F^{\mu\lambda}\bigg]\nonumber\\+\frac{\theta}{6}\epsilon^{\mu\nu\alpha\beta}H_{\mu\alpha\beta}+\frac{\theta M^2}{4}\epsilon^{\mu\nu\alpha\beta}\frac{n_\mu n^\lambda}{(n\cdot\partial)}H_{\lambda\alpha\beta}=0,
\end{align}
and
\begin{align}
\label{krm2}
\frac{\theta^2}{12m^2}\bigg[\partial_\lambda
H^{\lambda\mu\nu}+m_{B}^2\frac{n_\lambda}{(n\cdot\partial)}H^{\lambda\mu\nu}\nonumber\\+\frac{m_{B}^2}{2}\bigg(\frac{n^\mu n_\rho}{(n\cdot\partial)^2}\partial_\lambda H^{\rho\nu\lambda}+\frac{n^\nu n_\rho}{(n\cdot\partial)^2}\partial_\lambda H^{\rho\lambda\mu}\bigg)\bigg]\nonumber\\+\theta\epsilon^{\mu\nu\alpha\beta}F_{\alpha\beta}+\frac{\theta M^2}{2}\epsilon^{\mu\nu\alpha\beta}\frac{n_\alpha n^\lambda}{(n\cdot\partial)^2} F_{\lambda\beta}=0. 
\end{align}

To proceed further, let us introduce the dual fields associated with the field strength tensors $H^{\lambda\mu\nu}$ and $F^{\mu\nu}$ in order to establish a relation between the self-dual model and the $SIM(2)$ Maxwell-Kalb-Ramond model. Then, we define 
\begin{align}
\label{dC} C_\mu=-\frac{\theta}{6m^2}\epsilon_{\mu\nu\alpha\beta}H^{\nu\alpha\beta},\\
\label{dG} G_{\mu\nu}=\frac{\theta}{2}\epsilon_{\mu\nu\alpha\beta}F^{\alpha\beta}.   
\end{align}

In terms of $C_\mu$ and $G_{\mu\nu}$, we rewrite the equations of motion (\ref{krm1}) and (\ref{krm2}) as follows
\begin{align}
\label{ax1} G_{\mu\nu} - \frac{M^2}{8}\bigg(\frac{n_\mu n^\lambda}{(n\cdot\partial)^2}G_{\lambda\nu}-\frac{n_\nu n^\lambda}{(n\cdot\partial)^2}G_{\lambda\mu}\bigg)-\theta\epsilon_{\mu\nu\alpha\beta}\partial^\alpha C^{\beta}\nonumber\\+2\theta m_{B}^2 \epsilon_{\mu\nu\alpha\beta}\partial^\alpha \frac{n^\beta n^\lambda}{(n\cdot\partial)^2} C_\lambda-\frac{1}{2}\theta m_{B}^2 \epsilon_{\mu\nu\alpha\beta}\frac{n^\alpha}{n\cdot\partial}C^\beta=0,
 \end{align}
and
\begin{align}
\label{ax2} C_\mu-M^2\frac{n_\mu n^\lambda}{(n\cdot\partial)^2}C_\lambda-\frac{\theta}{2m^2}\epsilon_{\mu\nu\alpha\beta}\partial^\nu G^{\alpha\beta}\nonumber\\-\frac{\theta m_{A}^2}{4m^2}\epsilon_{\mu\nu\alpha\beta}\frac{n^\nu}{n\cdot\partial}G^{\alpha\beta}+\frac{\theta m_{A}^2}{2m^2}\epsilon_{\mu\nu\alpha\beta}\partial^\nu\frac{n^\alpha n_\lambda}{(n\cdot\partial)^2}G^{\lambda\beta}=0.    
\end{align}

By directly comparing the pairs of equations (\ref{ebf1}), (\ref{ebf2}) and (\ref{ax1}), (\ref{ax2}), we readily verify that the dual fields satisfy the following relations:
\begin{align}
\label{A} A_\mu\rightarrow C_\mu-M^2\frac{n_\mu n^\lambda}{(n\cdot\partial)^2}C_\lambda,\\
\label{B} B_{\mu\nu}\rightarrow G_{\mu\nu} - \frac{M^2}{8}\bigg(\frac{n_\mu n^\lambda}{(n\cdot\partial)^2}G_{\lambda\nu}-\frac{n_\nu n^\lambda}{(n\cdot\partial)^2}G_{\lambda\mu}\bigg)
\end{align}

With the above results, we have established the duality between the models at the classical level through the equations of motion in the free-field case. We find that the vector field of the VSR self-dual model is related to the dual field in Eq.~(\ref{dC}) of the MKR model, while the antisymmetric tensor field is related to the dual field in Eq.~(\ref{dG}). Both fields exhibit nonlocal VSR corrections. In the sequel, this equivalence will be revisited using the master Lagrangian approach.

%\textcolor{red}{\bf Stop here}
%%%%%%%%%%%%%%%%%%%%%%%%%%%%%%%%%%%%%%%%%%%%%%%%%%%%%%%%%%%%%%%%%%%%%%%%%%%%%%%%%%%%
\section{Duality via Master Lagrangian method}\label{s4}

 Previously, the connection between the SD and MKR models was established within the framework of very special relativity at the level of the equations of motion. The next step is to employ the master Lagrangian method to verify the interpolation between these models. First, we introduce two auxiliary fields, thereby transforming the Lagrangian density $\mathcal{L}_{MKR}$ into a first-order derivative form \cite{Maluf:2020fch}:
\begin{align}\label{master1}
\mathcal{L}_{M}=a\epsilon^{\mu\nu\alpha\beta}\Lambda_{\mu\nu}\partial_\alpha A_\beta+\frac{a\, M^2}{2}\epsilon^{\mu\nu\alpha\beta}\Lambda_{\mu\nu}\frac{n_\alpha}{(n\cdot\partial)} A_\beta+\frac{a\, m_{A}^2}{2}\epsilon^{\mu\nu\alpha\beta}\Lambda_{\mu\nu}\frac{n_\alpha}{(n\cdot\partial)^2}\partial_\beta (n\cdot A)\nonumber\\+b\epsilon^{\mu\nu\alpha\beta}\Pi_\mu\partial_\nu B_{\alpha\beta}+\frac{b\, M^2}{4}\epsilon^{\mu\nu\alpha\beta}\,\Pi_\mu\frac{n_\nu}{(n\cdot\partial)} B_{\alpha\beta}+\frac{b\, m_{B}^2}{2}\epsilon^{\mu\nu\alpha\beta}\,\Pi_\mu\partial_\nu\bigg[\frac{n_\alpha n^\lambda}{(n\cdot\partial)^2}B_{\lambda\beta}\bigg]\nonumber\\+c\,\Pi_\mu \Pi^\mu+d\,\Lambda_{\mu\nu}\Lambda^{\mu\nu}-\frac{\theta}{8} \epsilon^{\mu\nu\alpha\beta}F_{\mu\nu}  B_{\alpha\beta}-\frac{\theta m_{B}^2}{8} \epsilon^{\mu\nu\alpha\beta}\frac{n_\mu n^\lambda}{(n\cdot\partial)^2}F_{\lambda\nu}  B_{\alpha\beta}\nonumber\\-\frac{\theta}{12} \epsilon^{\mu\nu\alpha\beta} A_\mu H_{\nu\alpha\beta}-\frac{\theta}{8} \epsilon^{\mu\nu\alpha\beta} A_\mu \frac{n_\nu n^\lambda}{(n\cdot\partial)^2} H_{\lambda\alpha\beta},
\end{align}
where we must determine the constant coefficients $a$, $b$, $c$ and $d$. It is easy to observe that the master Lagrangian density is written in terms of the fields $A_{\mu}$, $B_{\mu\nu}$, $\Pi_{\mu}$ and $\Lambda_{\mu\nu}$. On the other hand, the lack of derivative terms and the presence of the mass term for both $\Pi_{\mu}$  and $\Lambda_{\mu\nu}$ guarantees the auxiliary nature of these fields.

First, we vary the action associated with $\mathcal{L}_{M}$, i.e., $\int d^4x \mathcal{L}_{M}$, with respect to both $\Pi_\mu$ and $\Lambda_{\mu\nu}$, obtaining the following equations of motion for the auxiliary fields:
\begin{align}
\label{Pf} \Pi^\mu=-\frac{a}{6c}\epsilon^{\mu\nu\alpha\beta}\bigg[3\partial_\nu B_{\alpha\beta}+\frac{3}{4}M^2\frac{n_\nu}{(n\cdot\partial)}B_{\alpha\beta}+3m_{B}^2\frac{n_\alpha}{(n\cdot\partial)^2}\partial_\nu(n^\lambda B_{\beta\lambda})\bigg],\\
\label{Lf} \Lambda^{\mu\nu}=-\frac{b}{4d}\epsilon^{\mu\nu\alpha\beta}\bigg[2\partial_\alpha A_\beta+\frac{1}{2}M^2\frac{n_\alpha}{(n\cdot\partial)}A_{\beta}+m_{A}^2\frac{n_\alpha}{(n\cdot\partial)^2}\partial_\beta(n\cdot A)\bigg].
\end{align}

Similarly, we can apply the above procedure to the fields $A_{\mu}$ and $B_{\mu\nu}$, allowing us to solve their equations of motion and obtain the following solution:
\begin{align}
\label{Af}   A_\mu+\frac{m_{A}^2}{2}\frac{n_\mu}{(n\cdot\partial)^2}(n\cdot A)=-\frac{a}{\theta}\Pi_\mu+\partial_\mu \Xi +\frac{m_A^2}{2}\frac{n_\mu}{(n\cdot \partial)} \Xi,\\
\label{Bf} B_{\mu\nu}+\frac{m_{B}^2}{2}\frac{n^\lambda}{(n\cdot\partial)^2}(n_\mu B_{\lambda\nu}+n_\nu B_{\lambda\mu})=\nonumber\\\frac{2b}{\theta}\Lambda_{\mu\nu}+\partial_\mu \Sigma_\nu-\partial_\mu \Sigma_\nu+\frac{m_B^2}{2}\bigg(\frac{n_\mu}{n\cdot\partial}\Sigma_\nu-\frac{n_\nu}{n\cdot\partial}\Sigma_\mu\bigg),
\end{align}
where $\Xi$ is an arbitrary scalar field and $\Omega_\mu$ is an arbitrary vector field. If we substitute (\ref{Af}) and (\ref{Bf}) into (\ref{master1}) and impose $\mathcal{L}_{M}=\mathcal{L}_{MKR}$, we arrive at the following relations
\begin{align}
\frac{a^2}{c}=\frac{\theta^2}{2m},\\
\frac{b^2}{d}=-\theta^2.
\end{align}
On the other hand, by substituting Eqs.~(\ref{Pf}) and (\ref{Lf}) into Eq.~(\ref{master1}), we recover the VSR self-dual Lagrangian $\mathcal{L}_{SD}$ with 
\begin{align}
c=\frac{m^2}{2},\\
d=-\frac{1}{4}.
\end{align}
With these relations, we can fix $a=b=\theta/2$ in Eq. (\ref{master1}), so that the master Lagrangian can be written in its final form:
\begin{align}\label{master2}
\mathcal{L}_{M}=\frac{\theta}{2}\epsilon^{\mu\nu\alpha\beta}\Lambda_{\mu\nu}\partial_\alpha A_\beta+\frac{\theta\, M^2}{8}\epsilon^{\mu\nu\alpha\beta}\Lambda_{\mu\nu}\frac{n_\alpha}{(n\cdot\partial)} A_\beta+\frac{\theta\, m_{A}^2}{4}\epsilon^{\mu\nu\alpha\beta}\Lambda_{\mu\nu}\frac{n_\beta}{(n\cdot\partial)^2}\partial_\alpha (n\cdot A)\nonumber\\+\frac{\theta}{2}\epsilon^{\mu\nu\alpha\beta}\Pi_\mu\partial_\nu B_{\alpha\beta}+\frac{\theta\, M^2}{8}\epsilon^{\mu\nu\alpha\beta}\,\Pi_\mu\frac{n_\nu}{(n\cdot\partial)} B_{\alpha\beta}+\frac{\theta\, m_{B}^2}{2}\epsilon^{\mu\nu\alpha\beta}\,\Pi_\mu\partial_\nu\bigg[\frac{n_\alpha n^\lambda}{(n\cdot\partial)^2}B_{\lambda\beta}\bigg]\nonumber\\+\frac{m^2}{2}\,\Pi_\mu \Pi^\mu-\frac{1}{4}\,\Lambda_{\mu\nu}\Lambda^{\mu\nu}-\frac{\theta}{8} \epsilon^{\mu\nu\alpha\beta}F_{\mu\nu}  B_{\alpha\beta}-\frac{\theta m_{B}^2}{8} \epsilon^{\mu\nu\alpha\beta}\frac{n_\mu n^\lambda}{(n\cdot\partial)^2}F_{\lambda\nu}  B_{\alpha\beta}\nonumber\\-\frac{\theta}{12} \epsilon^{\mu\nu\alpha\beta} A_\mu H_{\nu\alpha\beta}-\frac{\theta}{8} \epsilon^{\mu\nu\alpha\beta} A_\mu \frac{n_\nu n^\lambda}{(n\cdot\partial)^2} H_{\lambda\alpha\beta}. 
\end{align}
%Above, we have introduced the parameter $\chi=\pm 1$, which is related either to the self-duality ($+$) or to anti-self-duality ($-$) of the theory.

On the other hand, we can rewrite the equations of motion, Eqs.~(\ref{Af}) and (\ref{Bf}), using $C_\mu=-\frac{\theta}{6m^2}\epsilon_{\mu\nu\alpha\beta}H^{\nu\alpha\beta}$ and $G_{\mu\nu}=\frac{\theta}{2}\epsilon_{\mu\nu\alpha\beta}F^{\alpha\beta}$, as follows 
\begin{align}
\Pi_\mu\rightarrow C_\mu-M^2\frac{n_\mu n^\lambda}{(n\cdot\partial)^2}C_\lambda,\\
\Lambda_{\mu\nu}\rightarrow G_{\mu\nu} - \frac{M^2}{8}\bigg(\frac{n_\mu n^\lambda}{(n\cdot\partial)^2}G_{\lambda\nu}-\frac{n_\nu n^\lambda}{(n\cdot\partial)^2}G_{\lambda\mu}\bigg).    
\end{align}
We straightforwardly observe that the above set coincides with the relations previously obtained in Eqs.~(\ref{A}) and (\ref{B}) through a direct comparison of the equations of motion of the SD and MKR models. As discussed earlier, the master Lagrangian in Eq.~(\ref{master2}) generates both the $SIM(2)$ MKR and SD models within the VSR framework. Moreover, it is readily verified that $\mathcal{L}_{M}$ is gauge invariant under the transformations $\delta A_\mu=\tilde{\partial}_{\mu}\Lambda$ and $\delta B_{\mu\nu}=\tilde{\partial}_{\mu}\beta_\nu-\tilde{\partial}_{\nu}\beta_\mu$ with $\delta\Pi_{\mu}=\delta\Lambda_{\mu\nu}=0$. Additionally, the mechanism demonstrated in this section can be extended to a more general theory coupled to matter fields, as we will discuss in the next section.

%%%%%%%%%%%%%%%%%%%%%%%%%%%%%%%%%%%%%%%%%%%%%%%%%%%%%%%%%%%%%
\section{Coupling with fermionic matter}\label{s5}

\subsection{Gauge sector}

Having examined the duality between the free $SIM(2)$ SD and MKR models in the VSR framework using the master Lagrangian approach, we are now in a position to apply this procedure to analyze the coupling to matter fields. To this end, we define the VSR self-dual model linearly coupled to a fermionic field as follows:
\begin{align}
\mathcal{L}_{\psi-SD}=\mathcal{L}_{SD}+\Pi_\mu J^\mu+\Lambda_{\mu\nu}\mathcal{J}^{\mu\nu}+\mathcal{L}(\psi),
\end{align}
where $\mathcal{L}(\psi)$ denotes a generic VSR Lagrangian for the free Dirac field, while $J^\mu$ and $\mathcal{J}^{\mu\nu}$ are matter currents associated with $\Pi_\mu$ and $\Lambda_{\mu\nu}$, respectively. 

It is worth mentioning that these currents encompass both the usual contributions and the modifications induced by VSR effects. Moreover, they are constructed solely from Dirac fields and, at the linear level, do not depend on the gauge or self-dual fields. Beyond this approximation, a perturbative analysis based on Feynman diagrams would be required; however, such an investigation is left for a forthcoming paper.

Therefore, following the duality at the linear approximation, we consider the master Lagrangian (\ref{master2}) with linear coupling to the self-dual field:
\begin{align}\label{masterM}
\mathcal{L}_{M}=\frac{\theta}{2}\epsilon^{\mu\nu\alpha\beta}\Lambda_{\mu\nu}\partial_\alpha A_\beta+\frac{\theta\, M^2}{8}\epsilon^{\mu\nu\alpha\beta}\Lambda_{\mu\nu}\frac{n_\alpha}{(n\cdot\partial)} A_\beta+\frac{\theta\, m_{A}^2}{4}\epsilon^{\mu\nu\alpha\beta}\Lambda_{\mu\nu}\frac{n_\beta}{(n\cdot\partial)^2}\partial_\alpha (n\cdot A)\nonumber\\+\frac{\theta}{2}\epsilon^{\mu\nu\alpha\beta}\Pi_\mu\partial_\nu B_{\alpha\beta}+\frac{\theta\, M^2}{8}\epsilon^{\mu\nu\alpha\beta}\,\Pi_\mu\frac{n_\nu}{(n\cdot\partial)} B_{\alpha\beta}+\frac{\theta\, m_{B}^2}{2}\epsilon^{\mu\nu\alpha\beta}\,\Pi_\mu\partial_\nu\bigg[\frac{n_\alpha n^\lambda}{(n\cdot\partial)^2}B_{\lambda\beta}\bigg]\nonumber\\+\frac{m^2}{2}\,\Pi_\mu \Pi^\mu-\frac{1}{4}\,\Lambda_{\mu\nu}\Lambda^{\mu\nu}-\frac{\theta}{8} \epsilon^{\mu\nu\alpha\beta}F_{\mu\nu}  B_{\alpha\beta}-\frac{\theta m_{B}^2}{8} \epsilon^{\mu\nu\alpha\beta}\frac{n_\mu n^\lambda}{(n\cdot\partial)^2}F_{\lambda\nu}  B_{\alpha\beta}\nonumber\\-\frac{\theta}{12} \epsilon^{\mu\nu\alpha\beta} A_\mu H_{\nu\alpha\beta}-\frac{\theta}{8} \epsilon^{\mu\nu\alpha\beta} A_\mu \frac{n_\nu n^\lambda}{(n\cdot\partial)^2} H_{\lambda\alpha\beta} \nonumber\\+\Pi_\mu J^\mu+\Lambda_{\mu\nu}\mathcal{J}^\mu+\mathcal{L}(\psi).
\end{align}

By varying the action associated with $\mathcal{L}_M$ (\ref{masterM}) with respect to $A_\mu$ and $B_{\mu\nu}$, we obtain the following equations of motion:
\begin{align}
\label{Pfm} \Pi^\mu=-\frac{\theta}{2m^2}\epsilon^{\mu\nu\alpha\beta}\bigg[\partial_\nu B_{\alpha\beta}+\frac{1}{4}M^2\frac{n_\nu}{(n\cdot\partial)}B_{\alpha\beta}+m_{B}^2\frac{n_\alpha}{(n\cdot\partial)^2}\partial_\nu(n^\lambda B_{\beta\lambda})\bigg]-\frac{1}{m^2}J^\mu\\
\label{Lfm} \Lambda^{\mu\nu}=\theta\epsilon^{\mu\nu\alpha\beta}\bigg[\partial_\alpha A_\beta+\frac{1}{4}M^2\frac{n_\nu}{(n\cdot\partial)}A_{\beta}+\frac{1}{2}m_{A}^2\frac{n_\alpha}{(n\cdot\partial)^2}\partial_\beta(n\cdot A)\bigg]+2\mathcal{J}^{\mu\nu}.
\end{align}
Substituting Eqs. (\ref{Pfm}) and (\ref{Lfm}) into Eq. (\ref{masterM}), we obtain the relation
\begin{align}\label{sdf}
\mathcal{L}_{\psi-SD}=\mathcal{L}_{SD}+\Pi_\mu J^\mu+\Lambda_{\mu\nu}\mathcal{J}^{\mu\nu}+\mathcal{L}(\psi).
\end{align}
Additionally, substituting Eqs. (\ref{Af}) and (\ref{Bf}) into Eq. (\ref{masterM}), we obtain the result
\begin{align}\label{mkrf}
\mathcal{L}_{\psi-MKR}=\mathcal{L}_{MKR}-\theta\bigg(\epsilon^{\mu\nu\alpha\beta}\partial_\nu A_\mu+\frac{m_{A}^2}{2}\epsilon^{\mu\nu\alpha\beta}\frac{n_\rho}{(n\cdot \partial)}\partial_\nu (n\cdot A)+\frac{M^2}{4}\epsilon^{\mu\nu\alpha\beta}\frac{n_\mu}{(n\cdot \partial)} A_\rho\bigg) \mathcal{J}_{\alpha\beta}\nonumber\\-\frac{\theta}{2m^2}\bigg(\epsilon^{\mu\nu\alpha\beta} B_{\mu\nu}\, \partial_\alpha+\frac{m_{A}^2}{2}\epsilon^{\mu\nu\alpha\beta}B_{\mu\nu}\frac{n_\alpha}{(n\cdot \partial)}+\frac{M^2}{4}\epsilon^{\mu\nu\alpha\beta}\frac{n_\mu n^\lambda}{(n\cdot \partial)} B_{\lambda\nu}\,\partial_\alpha\bigg) J_\beta\nonumber\\-\frac{1}{2m^2} J_\mu J^\mu+ \mathcal{J}_\mu\mathcal{J}^\mu+\mathcal{L}(\psi). 
\end{align}

From the above Lagrangian, we observe that the MKR model acquires a Thirring-like term. This term is modified by VSR effects and involves only the matter fields. On the other hand, the relation between the models can be expressed through the following identification:
\begin{align}
\Pi_\mu\rightarrow C_\mu-M^2\frac{n_\mu n^\lambda}{(n\cdot\partial)^2}C_\lambda-\frac{1}{m^2}J_\mu,\\
\Lambda_{\mu\nu}\rightarrow G_{\mu\nu} - \frac{M^2}{8}\bigg(\frac{n_\mu n^\lambda}{(n\cdot\partial)^2}G_{\lambda\nu}-\frac{n_\nu n^\lambda}{(n\cdot\partial)^2}G_{\lambda\mu}\bigg)+2\mathcal{J}_{\mu\nu}.    
\end{align}

In contrast to the free-field case, the identifications between the fields $\Pi^{\mu}$ and $F^{\mu}$, as well as between the fields $\Lambda^{\mu\nu}$ and $G^{\mu\nu}$, are modified by the presence of current terms. These contributions will be relevant for the subsequent analysis of the fermionic sector in both models.

%%%%%%%%%%%%%%%%%%%%%%%%%%%%%%%%%%%%%%%
\subsection{Matter sector}

As shown previously, the duality between the MKR and SD models was established in the presence of a fermionic matter source. However, up to this point, we have focused exclusively on the gauge sector. The next step is therefore to verify the duality in the fermionic sector. By performing a functional variation of (\ref{sdf}), we get
\begin{align}\label{f1}
\frac{\delta}{\delta\psi}\int d^4x\, \mathcal{L}_{\psi-SD}=0\Rightarrow\frac{\delta \mathcal{L}(\psi)}{\delta\psi}=-\Pi_\mu\frac{\delta}{\delta\psi} J^\mu-\Lambda_{\mu\nu}\frac{\delta}{\delta\psi} \mathcal{J}^{\mu\nu}.
\end{align}
On the other hand, the gauge sector for the SD model with matter source reads 
\begin{align}
\label{ebf11} m^2 \Pi^\mu+\frac{\theta}{2}\epsilon^{\mu\nu\alpha\beta}\partial_\nu \Lambda_{\alpha\beta}+\frac{\theta M^2}{8}\epsilon^{\mu\nu\alpha\beta}\frac{n_\nu}{(n\cdot\partial)}\Lambda_{\alpha\beta}=-J^\mu,\\
\label{ebf22}    -\frac{1}{2}\Lambda^{\mu\nu}+\frac{\theta}{2}\epsilon^{\mu\nu\alpha\beta}\partial_\alpha \Pi_\beta +\frac{\theta M^2}{8}\epsilon^{\mu\nu\alpha\beta}\frac{n_\alpha}{(n\cdot\partial)}\Pi_\beta=-\mathcal{J}^{\mu\nu}.
\end{align}
The above set of equations obey the following constraints
\begin{align}
    m^2\partial_\mu \Pi^\mu+\frac{M^2}{4}\frac{n_\mu}{(n\cdot\partial)} \Pi^\mu=-\partial_\mu J^\mu-\frac{M^2}{4}\frac{n_\mu}{(n\cdot\partial)} J^\mu,\\
     m^2\partial_\mu \Lambda^{\mu\nu}+\frac{M^2}{4}\frac{n_\mu}{(n\cdot\partial)}\Lambda^{\mu\nu}=2\partial_\mu \mathcal{J}^{\mu\nu}+\frac{M^2}{2}\frac{n_\mu}{(n\cdot\partial)} \mathcal{J}^{\mu\nu}.
\end{align}
With the help of these constraints, we can express $\Pi_{\mu}$ and $\Lambda_{\mu\nu}$ separately, after some algebraic manipulations, as follows:
\begin{align}
\label{pf}    \Pi_{\mu}=-\frac{\hat{R}}{\theta^2}\bigg[J_\mu+\frac{\theta^2}{m^2}\bigg(\partial_\mu\partial^\rho J_\rho+\frac{M^2}{4}\frac{n^\rho\partial_\mu+n_\mu\partial^\rho}{(n\cdot\partial)}J_\rho+\frac{M^4}{16}\frac{n_\mu n^\rho}{(n\cdot\partial)^2}J_\rho\bigg)\nonumber\\+\theta\epsilon_{\mu\nu\alpha\beta}\partial^\nu\mathcal{J}^{\alpha\beta}+\frac{\theta M^2}{4}\epsilon_{\mu\nu\alpha\beta}\frac{n^\nu}{(n\cdot\partial)}\mathcal{J}^{\alpha\beta}\bigg],
    \end{align}
and
\begin{align}
\label{lf} \Lambda_{\mu\nu}=-\frac{\hat{R}}{\theta^2}\bigg[-2m^2\mathcal{J}_{\mu\nu}+2\theta^2\partial^\rho(\partial_\mu\mathcal{J}_{\nu\rho}-\partial_\nu\mathcal{J}_{\mu\rho})+\frac{M^2}{4}\frac{1}{(n\cdot\partial)}[(n_\mu\partial^\rho+n^\rho\partial_\mu)\mathcal{J}_{\nu\rho}\nonumber\\-(n_\nu\partial^\rho+n^\rho\partial_\nu)\mathcal{J}_{\mu\rho}]+\frac{M^4}{16}\frac{n^\rho}{(n\cdot\partial)}(n_\mu\mathcal{J}_{\nu\rho}-n_\nu\mathcal{J}_{\mu\rho})\nonumber\\+\theta\epsilon_{\mu\nu\alpha\beta}\partial^\alpha J^\beta+\frac{\theta M^2}{4}\epsilon_{\mu\nu\alpha\beta}\frac{n^\alpha}{(n\cdot\partial)}J^\beta\bigg],   
\end{align}
where we have defined the wave operator $\hat{R}$ as being
\begin{align}
 \hat{R}=\frac{1}{\square+\frac{M^2}{2}+\frac{m^2}{\theta^2}}.
\end{align}

Let us proceed by rewriting Eq.~(\ref{f1}) solely in terms of the current contributions. To this end, we substitute the previously obtained expressions for $\Pi_{\mu}$ and $\Lambda_{\mu\nu}$, which leads to
\begin{align}
\frac{\delta \mathcal{L}(\psi)}{\delta\psi}=\frac{\hat{R}}{\theta^2}\bigg[J_\mu+\frac{\theta^2}{m^2}\bigg(\partial_\mu\partial^\rho J_\rho+\frac{M^2}{4}\frac{n^\rho\partial_\mu+n_\mu\partial^\rho}{(n\cdot\partial)}J_\rho+\frac{M^4}{16}\frac{n_\mu n^\rho}{(n\cdot\partial)^2}J_\rho\bigg)\nonumber\\+\theta\epsilon_{\mu\nu\alpha\beta}\partial^\nu\mathcal{J}^{\alpha\beta}+\frac{\theta M^2}{4}\epsilon_{\mu\nu\alpha\beta}\frac{n^\nu}{(n\cdot\partial)}\mathcal{J}^{\alpha\beta}\bigg]\frac{\delta}{\delta\psi} J^\mu\nonumber\\+\frac{\hat{R}}{\theta^2}\bigg[-2m^2\mathcal{J}_{\mu\nu}+2\theta^2\partial^\rho(\partial_\mu\mathcal{J}_{\nu\rho}-\partial_\nu\mathcal{J}_{\mu\rho})+\frac{M^2}{4}\frac{1}{(n\cdot\partial)}[(n_\mu\partial^\rho+n^\rho\partial_\mu)\mathcal{J}_{\nu\rho}\nonumber\\-(n_\nu\partial^\rho+n^\rho\partial_\nu)\mathcal{J}_{\mu\rho}]+\frac{M^4}{16}\frac{n^\rho}{(n\cdot\partial)}(n_\mu\mathcal{J}_{\nu\rho}-n_\nu\mathcal{J}_{\mu\rho})\nonumber\\+\theta\epsilon_{\mu\nu\alpha\beta}\partial^\alpha J^\beta+\frac{\theta M^2}{4}\epsilon_{\mu\nu\alpha\beta}\frac{n^\alpha}{(n\cdot\partial)}J^\beta\bigg]\frac{\delta}{\delta\psi} \mathcal{J}^{\mu\nu}. 
\end{align}
The above result is a nonlocal differential equation written solely in terms of the matter fields. We point out that the result obtained in Ref. \cite{Maluf:2020fch} can be recovered in the case of $m_{A}=m_{B}=0$. We now turn our attention to the MKR model. Accordingly, by varying the action (\ref{mkrf}), we obtain
\begin{align}\label{efm1}
\frac{\delta}{\delta\psi}\int d^4x\, \mathcal{L}_{\psi-MKR}=0\Rightarrow\frac{\delta \mathcal{L}(\psi)}{\delta\psi}=-\bigg(C_\mu-M^2\frac{n_\mu n^\lambda}{(n\cdot\partial)^2}C_\lambda-\frac{1}{m^2}J_\mu\bigg)\frac{\delta}{\delta\psi} J^\mu\nonumber\\-\bigg[G_{\mu\nu} - \frac{M^2}{8}\bigg(\frac{n_\mu n^\lambda}{(n\cdot\partial)^2}G_{\lambda\nu}-\frac{n_\nu n^\lambda}{(n\cdot\partial)^2}G_{\lambda\mu}\bigg)+2\mathcal{J}_{\mu\nu}\bigg]\frac{\delta}{\delta\psi} \mathcal{J}^{\mu\nu}.
\end{align}
By substituting Eqs.~(\ref{pf}) and (\ref{lf}) into the above equation, we obtain
\begin{align}
\frac{\delta \mathcal{L}(\psi)}{\delta\psi}=-\Pi_\mu\frac{\delta}{\delta\psi} J^\mu-\Lambda_{\mu\nu}\frac{\delta}{\delta\psi} \mathcal{J}^{\mu\nu}.
\end{align}
Finally, we arrive at
\begin{align}\label{efm2}
\frac{\delta \mathcal{L}(\psi)}{\delta\psi}=\frac{\hat{R}}{\theta^2}\bigg[J_\mu+\frac{\theta^2}{m^2}\bigg(\partial_\mu\partial^\rho J_\rho+\frac{M^2}{4}\frac{n^\rho\partial_\mu+n_\mu\partial^\rho}{(n\cdot\partial)}J_\rho+\frac{M^4}{16}\frac{n_\mu n^\rho}{(n\cdot\partial)^2}J_\rho\bigg)\nonumber\\+\theta\epsilon_{\mu\nu\alpha\beta}\partial^\nu\mathcal{J}^{\alpha\beta}+\frac{\theta M^2}{4}\epsilon_{\mu\nu\alpha\beta}\frac{n^\nu}{(n\cdot\partial)}\mathcal{J}^{\alpha\beta}\bigg]\frac{\delta}{\delta\psi} J^\mu\nonumber\\+\frac{\hat{R}}{\theta^2}\bigg[-2m^2\mathcal{J}_{\mu\nu}+2\theta^2\partial^\rho(\partial_\mu\mathcal{J}_{\nu\rho}-\partial_\nu\mathcal{J}_{\mu\rho})+\frac{M^2}{4}\frac{1}{(n\cdot\partial)}[(n_\mu\partial^\rho+n^\rho\partial_\mu)\mathcal{J}_{\nu\rho}\nonumber\\-(n_\nu\partial^\rho+n^\rho\partial_\nu)\mathcal{J}_{\mu\rho}]+\frac{M^4}{16}\frac{n^\rho}{(n\cdot\partial)}(n_\mu\mathcal{J}_{\nu\rho}-n_\nu\mathcal{J}_{\mu\rho})\nonumber\\+\theta\epsilon_{\mu\nu\alpha\beta}\partial^\alpha J^\beta+\frac{\theta M^2}{4}\epsilon_{\mu\nu\alpha\beta}\frac{n^\alpha}{(n\cdot\partial)}J^\beta\bigg]\frac{\delta}{\delta\psi} \mathcal{J}^{\mu\nu}.
\end{align}

In this way, we show that the SD fermionic sector described by Eq.~(\ref{efm1}) is equivalent to the MKR fermionic sector described by Eq.~(\ref{efm2}). Consequently, we establish the classical duality between the MKR and SD models coupled to fermionic matter in the VSR framework. Moreover, we emphasize that, in the absence of VSR effects, we recover the known results for the duality in the fermionic sector previously investigated in Refs.~\cite{Menezes:2002nj, Maluf:2020fch}.

%%%%%%%%%%%%%%%%%%%%%%%%%%%%%%%%%%%%%%%%%%%%%%%%%%%%%%%%%%%%%%%%%%%%%
\section{Final remarks}\label{s6}

We have revisited and extended the analysis of the classical duality between the self-dual (SD) and topologically massive models involving the $B \wedge F$ term in $(3+1)$-dimensional spacetime within the framework of Very Special Relativity (VSR). We have demonstrated that the presence of VSR induces nonlocal modifications in both the Maxwell–Kalb–Ramond (MKR) and self-dual models, thereby altering the structure of the standard Lorentz-invariant duality.

Initially, the correspondence between the $SIM(2)$-SD and $SIM(2)$-MKR models was established at the classical level through a direct comparison of their equations of motion. We verified that the presence of nonlocal $SIM(2)$ operators modifies the dual mapping between the vector and antisymmetric tensor fields, while preserving the essential equivalence between the two models. Subsequently, we employed the master Lagrangian approach to examine this duality. This method provides a systematic and elegant framework to establish the equivalence and to demonstrate the gauge-invariant interpolation between both formulations.

Furthermore, we analyzed the duality in the presence of fermionic matter. By minimally coupling the self-dual model to Dirac fields, we found that Thirring-like current--current interactions naturally emerge in the MKR counterpart. These interactions, modified by nonlocal VSR contributions, preserve the equivalence between the bosonic and fermionic sectors of the two models. Moreover, this formalism elucidates how VSR effects deform the standard local duality by introducing nonlocal terms proportional to inverse derivatives along the preferred null direction defined by $n^\mu$.

Finally, our results generalize the well-known topologically massive dualities into the VSR framework. In particular, we extend our previous work \cite{Belchior:2024eyz} by promoting the $SIM(1)$ symmetry to $SIM(2)$ symmetry through the inclusion of the $B \wedge F$ term. Moreover, our findings highlight that VSR-invariant extensions can consistently support topological mass generation mechanisms without introducing additional dynamical degrees of freedom.

\section*{Acknowledgments}
The authors thank the Funda\c{c}\~{a}o Cearense de Apoio ao Desenvolvimento Cient\'{i}fico e Tecnol\'{o}gico (FUNCAP), the Coordena\c{c}\~{a}o de Aperfei\c{c}oamento de Pessoal de N\'{i}vel Superior (CAPES), and the Conselho Nacional de Desenvolvimento Cient\'{i}fico e Tecnol\'{o}gico (CNPq).  Fernando M. Belchior has been partially supported by CNPq with Grant No. 161092/2021-7. Robert V. Maluf thanks the CNPq for Grant no. 311393/2025-0.

\end{document}